\documentstyle[preprint,aps,eqsecnum]{revtex}
 
\def\beq{\begin{equation}}  
\def\beq#1{\begin{equation} \label{#1}}
\def\eeq{\end{equation}}
\newcommand{\bea}{\begin{eqnarray}}
\newcommand{\eea}{\end{eqnarray}}
\newcommand{\ber}{\begin{eqnarray}}
\newcommand{\eer}{\end{eqnarray}}
\newcommand{\bers}{\begin{eqnarray*}}
\newcommand{\eers}{\end{eqnarray*}}

\def\ket#1{\left\vert #1\right\rangle}

\begin{document}

\title{
Experimental Challenges for QCD - The past and the future}

\author{Harry J. Lipkin}

\address{Department of Particle Physics, 
        Weizmann Institute of Science, \\ 
        Rehovot, Israel \\
	E-mail: ftlipkin@clever.weizmann.ac.il}  
 \address{School of Physics and Astronomy,
Raymond and Beverly Sackler Faculty of Exact Sciences,\\
        Tel Aviv University, Tel Aviv, Israel }
      \address{High Energy Physics Division, Argonne National Laboratory,\\ 
Argonne, IL 60439-4815, USA\\ 
E-mail: lipkin@hep.anl.gov}

\maketitle
\begin{abstract} 
The past leaves  the surprising experimental successes of the simple constituent
quark model to be expained by QCD. 
Surprising agreement with experiment from simple Sakharov-Zeldovich model
(1966) having quarks with effective masses and hyperfine interaction. Nambu's
(1966) Colored quarks with gauge gluons gave mass spectrum with only $qqq$ and
$\bar q$ bound states.
The  future opens the way to new insight
into QCD from heavy flavor experiments.
\end{abstract} 
\section{Introduction}

QCD is supposed to explain 
everything about Hadron Physics - But How?

QED is supposed to explain everything about Superconductivity .

Will explaining Hadron Spectrocopy from QCD be as 
difficult as explaining Superconductivity from  QED? 

The Past leaves us with many experimental regularities which await explanation
by QCD. The future offers many new experimental opportunities to learn about QCD
from heavy flavvor physics. .

\subsection { {The Past - pre-QCD questions - Challenges for QCD}}

\subsubsection {What is a Hadron?}

Present attempts to describe hadrons recall the story of the blind men and the
elephant\cite{PBIGSKY} . Each investigation finds one particular property of
hadrons and many contradictory conclusions arise that are all correct,
\begin{enumerate}
\item   A pion is a
Goldstone Boson and a proton is a Skyrmion, 
\item A pion is two-thirds of a
proton. The simple quark model prediction $ \sigma_{tot}(\pi^-p) \approx
(2/3)\cdot \sigma_{tot}(pp) $ \cite{LevFran,LS} still fits experimental data
better than 7\% up to 310 Gev/c\cite{PAQMREV}; 
\item The $a_1$ is a $q \bar q$ pair in a $^3P_1$ state  similar
to other $^3P$ states:  scalar and  tensor ($a_2$)
\item The $a_1$ is the chiral partner of the $\rho$ coupled similarly to the 
$W$.
\item The $\eta$  and  $\eta'$ are orthogonal linear combinations of the same 
strange and nonstrange ground state wave functions
\item   The $\eta$  and  $\eta'$ contain other components like glueballs or 
radial excitations
\item Mesons and Baryons are made
of the same quarks. Describing both as simple composites of asymptotically free
quasiparticles with a unique effective mass value predicts hadron masses,
magnetic moments and hyperfine splittings\cite{SakhZel,ICHJLmass,HJLMASS}. 
\item Lattice QCD can give all the answers, 
\item Lattice calculations disagree on
whether the H dibaryon is bound and offer no hope of settling this question
until much bigger lattices are available\cite{PBIGSKY}. 
\end{enumerate}
\subsubsection {What is a good hadronic symmetry? Many contradictions}
 
\begin{enumerate}
\item Light (uds) SU(3)
symmetry and Heavy Quark symmetry (cbt) are good; 
\item Light (uds) SU(3)
symmetry is bad. All nontrivial hadron states violate SU(3). All light V, A and
T mesons have good isospin symmetry with flavor mixing in (u.d) space and no $s
\bar s$ component; e.g. $\rho, \omega$. 
\item The s-quark is a heavy quark.
Flavor mixing in mass eigenstates predicted by SU(3) is not there. Most
nontrivial strange hadron states satisfy (scb) heavy quark symmetry with no
flavor mixing.; e.g. $\phi, \psi, \Upsilon$. 
\item  Light (uds) SU(3) symmetry is the basis of  Cabibbo theory of weak
interactions and 
gives  excellent description of hyperon decays. 
\item Violation of the Gottfried sum rule shows the proton sea is not 
isoscalar. 
\item Isospin symmetry requires a proton with an isovector sea to have a
component with a valence neutron and a charged sea
\item SU(3) requires proton with isovector sea to have a
component with a valence hyperon and a strange sea to satisfy Cabibbo theory
for vector current.
\item Experiment shows SU(3) symmetry manifestly broken in proton sea 
\item No consistent explanation of Gottfried violation, strange sea
suppression breaking SU(3) and Cabibbo theory requiring good SU(3).
\item Why are the $\omega$ and $\rho$ degenerate while the $\eta$  
and $\pi$ are not? 
Is there a
symmetry beyond SU(3) that forbids octet-singlet splitting for vectors but not
for pseudoscalars??
\end{enumerate}
\subsubsection {Why do the constituent quark model and the OZI rule work so 
well?}

Surprising agreement with experiment from simple Sakharov-Zeldovich model
(1966) having quarks with effective masses and hyperfine interaction. Nambu's
(1966) Colored quarks with gauge gluons gave mass spectrum with only $qqq$ and
$\bar q$ bound states.

The topological quark-line OZI rule does not follow from any symmetry and
predicts  experiments successfully without any solid theoretical
justification.

\subsection { {The Future - Heavy Flavor Decays Give New Insight}}

\begin{enumerate} 

\item Weak Decays need hadron models and QCD to interpret decays, but have 	
too many diagrams, too many free parameters 

\item No rigorous QCD results for FSI and strong phases

\item Too many decay modes, too much data.
Need phenomenologists to choose data for analysis

\item Experimental results from B and Charm factories 
that defy conventional wisdom can provide clues to
new physics and inadequacies in hadron models. 

\end{enumerate}

\section{  {The Past -  The constituent quark model and other 
      pre-QCD Challenges for QCD}}

\subsection{The Sakharov-Zeldovich 1966 Quark model (SZ66)}

\subsubsection{The Model}   
Andrei Sakharov, a pioneer in quark-hadron physics asked in 1966
``Why are the $\Lambda$ and $\Sigma$ masses
different? They are made of the same quarks". Sakharov and 
Zeldovich\cite{SakhZel}. 
assumed a  quark model for hadrons with a flavor dependent linear mass
term and hyperfine interaction,
\begin{equation}
M = \sum_i m_i + \sum_{i>j} 
{{\vec{\sigma}_i\cdot\vec{\sigma}_j}\over{m_i\cdot m_j}}\cdot v^{hyp}_{ij} 
\end{equation}
where $m_i$ is the effective mass of quark $i$, $\vec{\sigma}_i$ is a quark 
spin operator and $v^{hyp}_{ij}$ is
a hyperfine interaction with different strengths
but the same flavor dependence
for $qq$ and $\bar q q$ interactions.

This model can be considered analogous to the BCS description of
superconductivity. The constituent quarks are quasiparticles of unknown
structure with a background of a condensate. They have effective masses not
simply related to the bare current quark masses, and somehow including all
effects of confinement and other flavor independent potentials. The only
contribution to hadron masses not already included is a flavor-dependent
two-body hyperfine interaction inversely proportional to the product of these
same effective quark masses. Hadron magnetic moments are described simply by
adding the contributions of the moments of these constituent quarks with Dirac
magnetic moments having a scale determined by the same  effective masses. The
model describes low-lying excitations of a complex system with remarkable
success. 

\subsubsection{Striking Results and Predictive Power}

Sakarov and Zeldovich already in 1966 obtained two relations between meson and
baryon masses in remarkable agreement with experiment.
Both the mass difference $m_s-m_u$ between strange and nonstrange quarks and
their mass ratio $m_s/m_u$ have the same values 
when calculated from baryon masses and meson masses\cite{SakhZel} 

\begin{equation}
\langle m_s-m_u \rangle_{Bar}= M_\Lambda-M_N=177\,{\rm MeV}
\end{equation}

\begin{equation}
\langle m_s-m_u \rangle_{mes} =
{{3(M_{K^{\scriptstyle *}}-M_\rho )
+M_K-M_\pi}\over 4} =180\,{\rm MeV}
\end{equation}

\begin{equation}
\left({{m_s}\over{m_u}}\right)_{Bar} =
{{M_\Delta - M_N}\over{M_{\Sigma^*} - M_\Sigma}} = 1.53 =
\left({{m_s}\over{m_u}}\right)_{Mes} =
{{M_\rho - M_\pi}\over{M_{K^*}-M_K}}= 1.61
\end{equation}

Further extension of this approach led to two more relations for $m_s-m_u$ when
calculated from baryon masses and meson masses\cite{ICHJLmass,HJLMASS}. and to
three magnetic moment predictions with no free parameters\cite{DGG,Protvino}
 
\begin{equation}
\langle m_s-m_u \rangle_{mes} =
{{3 M_\rho + M_\pi}\over 8}
\cdot
\left({{M_\rho - M_\pi}\over{M_{K^*}-M_K}} - 1 \right)
= 178\,{\rm MeV}.
\end{equation}
\begin{equation}
\langle m_s-m_u \rangle_{Bar}= 
{{M_N+M_\Delta}\over 6}\cdot
\left({{M_{\Delta}-M_N}\over
{M_{\Sigma^{\scriptstyle *}}-M_\Sigma}} - 1 \right)
=190\,{\rm MeV}.
\end{equation}
\begin{equation}\mu_\Lambda=
-0.61
\,{\rm n.m.}=\mu_\Lambda =
-{\mu_p\over 3}\cdot {{m_u}\over{m_s}} =
-{\mu_p\over 3} {{M_{\Sigma^*} - M_\Sigma} \over{M_\Delta - M_N}}
=-0.61 \,{\rm n.m.}
\end{equation}

\begin{equation}
-1.46 =
{\mu_p \over \mu_n} =
-{3 \over 2}
\end{equation}

\begin{equation}
\mu_p+\mu_n= 0.88 \,{\rm n.m.}
={M_{\scriptstyle p}\over 3m_u}
={2M_{\scriptstyle p}\over M_N+M_\Delta}=0.865 \,{\rm n.m.}
\end{equation}

Also in 1966 Levin and Frankfurt\cite{LevFran} noted a remarkable systematics
in hadron-nucleon total cross sections indicating that mesons and baryons were
made of the same basic building blocks. The analysis supporting their ratio of
3/2 between baryon-nucleon and nucleon-nucleon cross sections has been 
refined\cite{LS} 
and consistently confirmed by new experiments\cite{PAQMREV}.
QCD calculations have not yet explained such remarkably successful simple
constituent quark model results. A search for new experimental input to guide
us is therefore of interest. 

\subsection{{The A.....Z or OZI rule and QCD}} 

\subsubsection{OZI for light quarks}

No rigorous QCD derivation has yet been found for this flavor-topology rule 
arising also in duality  diagrams of Regge phenomenology where  leading
t-channel exchanges are dual to s-channel resonances and in more modern planar
quark diagrams in large $N_c$ QCD. It has been repeatedly confirmed in a large
variety of experimental results and theoretical analyses for strong interaction
three-point and four-point functions,  beginning with the first controversial
prediction  relating final states in completely different isospin and
flavor-SU(3) multiplets unrelated by any known symmetry.
\beq{OZI}
\sigma(K^-p \rightarrow \Lambda \rho^o)= \sigma(K^-p
\rightarrow \Lambda \omega)     
\eeq

Is this connected with the $\omega - \rho$ degeneracy?

\subsubsection{OZI for heavy flavors - Why is $J/\psi$ narrow?}

One diagram $ J/\psi \rightarrow 3G \rightarrow $ light hadrons fits the
narrowness but $ J/\psi \rightarrow D \bar D \rightarrow $ light hadrons is
larger and neglected  along with ad hoc forbidden  ``hairpin diagrams". Hand
waving explanations with cancellations  give predictive power which can be
tested with future experimental data and still challenge QCD for explanation.

\subsection{Problems with the $\eta$ and $\eta'$}

The pseudoscalars are conventionally decribed by adding an additional mass
contribution to the SU(3) singlet state, thus breaking U(3) while conserving
SU(3) and leaving SU(3) breaking as entirely due to quark mass differences.. The dynamical origin of this additional
singlet contribution is still unclear and controversial, with some models
attributing it  the annihilation of an $q \bar q$ pair into gluons or
instantons and no reason to limit the mixing to only ground state $q \bar
q$ wave functions. Admixtures of radial excitations and glueballs have been 
considered.

\subsection{How to go beyond SZ66 with QCD}

Many approaches are being investigated to use QCD in the description of hadron
spectroscopy\cite{hadproc}.
The complexity of QCD calculations necessitates the introduction of ad hoc
approximations and free parameters to obtain results, thus losing the simplicity
of the constituent quark model, with its ability to make many independent
predictions with very few parameters, There is also a tendency to lose some of
the good results of the constituent quark model; namely 
\begin{itemize}    
\item  The universal treatment of mesons and baryons made of the same quarks
\item  The spin dependence of hadron masses as a hyperfine interaction
\item  The appearance of the same effective quark masses in hadron masses, 
spin splittings and magnetic moments 
\item  The systematic regularities relating meson-nucleon and baryon-nucleon 
cross sections 
\end{itemize}

While none of these results can be considered to have a firm theoretical 
foundation based on QCD, it is difficult simply to dismiss the striking
agreement with experiment and the successful predictive power as purely 
purely accidental.

\section { {The Future - Heavy Flavor Physics Gives New Insight}}

Weak Decays need hadron models and QCD to interpret decays, but have 	
too many diagrams and too many free parameters.  
Use of flavor topology can simplify analyses on one hand and
challenge QCD to explain them if they work.

\subsection{ Experimental systematics challenging conventional wisdom}

\subsubsection{{Universality of vector dominance couplings}}

The large branching ratios observed\cite {PDG}  for the appearance of the
$a_1(1260)^{\pm}$ in all quasi-two-body decays $D \rightarrow  a_1(1260)^{\pm}
X$ and  $B \rightarrow  a_1(1260)^{\pm}  X$  are comparable to  those observed
for  $\pi^{\pm} X$ and  $\rho^{\pm} X$.  
No decays to the other p-wave mesons are
within an order of magnitude of these values; e.g  the  difference between the
$a_1$ and the $a_2$.  All 24 $B$ decays of the form $ B \rightarrow  \bar D
W^+\rightarrow  \bar D M^+ $ , where $M$ can denote  $a_1, \rho,  \pi,
{\ell}^+\nu_{\ell}, D_s, D^*_s $, are dominant with branching ratios above 
$0.3\%$. Other $B$-decay modes have upper  limits in the $10^{-4}$ ball park,
including 
the absence with significant upper limits of  neutral decays $B^o \rightarrow
\bar D^o M^o $ which are coupled by strong final state interactions to $B^o
\rightarrow D^- M^+ $.

These experimental systematics suggested a 
``vector-dominance" model\cite{vecdom} where the initial hadron state $i$
decays to a final state $f$ by emitting a $W^{\pm}$ which then hadronizes into
an $a_1^+$, $\rho^+$ or $\pi^+$, along  with a universality relation,

$$
[if\pi]\equiv {{BR[ i \rightarrow  f \pi^+]}\over{BR[i \rightarrow  f \rho^+]
}}\approx \left|{{W^+  \rightarrow  \pi^+ }\over{W^+  \rightarrow 
 \rho^+ }}\right|^2
$$ 
$$
[ifa]\equiv {{BR[ i \rightarrow  f  a_1(1260)^+]}\over{BR[i \rightarrow 
 f \rho^+]
}}\approx \left|{{W^+  \rightarrow  a_1^+ }\over{W^+  \rightarrow  
\rho^+ }}\right|^2
$$ 
for all states $i$  and $f$ with corrections for phase space. 
$$
[D^+   \bar K^o \pi ]\approx
[D^o  K^-  \pi]\approx
[B^o  D^-  \pi]\approx
[B^o D^{*-}  \pi]\approx
[B^+   \bar D^o  \pi]\approx
[B^+   \bar D^{*o}\pi]
$$
$$
 .44  \pm   .17   \approx 
  .35 \pm    .09 \approx 
   .38  \pm    .08  \approx   
 .41 \pm    .20    \approx 
  .40 \pm    .06  \approx  
   .30   \pm    .07   
$$  
$$
[D^+   \bar K^o  a]\approx
[D^o  K^-  a]\approx
[B^o  D^-  a]\approx
[B^o D^{*-}a]\approx
[B^+   \bar D^o  a]\approx
[B^+   \bar D^{*o}  a]
$$
$$
  1.2   \pm   .5  \approx   
   .68  \pm    .12   \approx  
   .8  \pm    .4 \approx    
  1.9   \pm   1.0   \approx   
   .37  \pm    .30  \approx    
  1.2   \pm    .4   
$$ 
 The $a_1$ data have large errors, b ut the experimental
ratios  $[ifa]$ are all consistent with 0.7,  and more than order of magnitude higher
than other upper limits 
$$
[D^o  K^-  a_2^+] < 0.019 \pm 0.002; ~ ~ ~ ~ 
[D^+   \bar K^o  a_2^+] < 0.045 \pm 0.017
$$ 
  That such widely different decays should agree so well is
impressive and suggests further investigation. e.g. reducing the experimental
errors and looking for more decay modes like 
$D^+_s \rightarrow \phi a_1$, $D^+_s \rightarrow \omega a_1$,
$D^+  \rightarrow K^{*0} a_1$ and
$D^0  \rightarrow K^{*-} a_1$.

\subsubsection {Vector-Dominance Decays of the $B_c$}

The $B_c$ meson is identified against a large combinatorial background by decay
modes including a $J/\psi$. 
Vector dominance decay modes including the $J/\psi$ are expected to have 
relatively large branching ratios. These include:
$J/\psi \rho^+ $, $J/\psi a_1^+ $,  $J/\psi \pi^+ $, $J/\psi D^*_s$,
$J/\psi D_{s1A}$, and $J/\psi D_s $. 
The corresponding modes with a $\psi'$ instead of a $J/\psi$ are expected to 
have comparable branching ratios.

\subsubsection{{Puzzles in Singly-Suppressed Charm Decays}}

Two Cabibbo suppressed $D^+$ decay modes have anomalously high branching ratios
which are not simply explained by any model\cite{nuclolip}.
\beq{w1}
BR[D^+ \rightarrow  K^*(892)^+\bar K^o] = 3.2 \pm 1.5\%
\eeq
\beq{w2}
BR[D^+ \rightarrow  K^*(892)^+\bar K^*(892)^o] = 2.6 \pm 1.1\%
\eeq
These are the same order as corresponding Cabibbo allowed branching ratios
\beq{a1}
BR[D^+ \rightarrow  \rho^+\bar K^o] = 6.6 \pm 2.5\%
\eeq 
\beq{a2}
BR[D^+ \rightarrow  \rho^+\bar K^*(892)^o] = 2.1 \pm 1.3\%
\eeq 
The dominant tree diagrams for these corresponding allowed and
suppressed decays differ only in the weak vertices $c \rightarrow W^+ + s
\rightarrow \rho^+ + s$ and $c \rightarrow W^+ + s
\rightarrow K^*(892)^+ + s$ and have the same hadronization of the strange quark
s and spectator $\bar d$. These diagrams should show the expected Cabibbo
suppression which is not observed.  

All standard model diagrams that
can contribute to these  anomalously 
enhanced decays (\ref{w1}-\ref{w2}) are related by symmetries to
a very similar diagrams for one of the following  decay modes which show the
expected Cabibbo suppression

\beq{s1}
BR[D^+ \rightarrow  K^+\bar K^*(892)^o] = 0.42 \pm 0.05\%
\eeq 
\beq{s2}
BR[D^o \rightarrow  K^*(892)^+K^-] = 0.35 \pm 0.08\%
\eeq 
\beq{s3}
BR[D^o \rightarrow  K^*(892)^-K^+] = 0.18 \pm 0.01\%
\eeq 
\beq{s4}
BR[D^o \rightarrow  K^*(892)^o\bar K^o] < 0.08 \%
\eeq 
\beq{s5}
BR[D^o \rightarrow \bar K^*(892)^oK^o] < 0.16 \%
\eeq 
\beq{s6}
BR[D^o \rightarrow  K^*(892)^o\bar K^*(892)^o] = 0.14 \pm 0.05\%
\eeq 
There is no simple diagram that enhances the suppressed diagrams
(\ref{w1}-\ref{w2}) without also enhancing others that show no experimental 
enhancement.   
It is therefore of interest to check the branching ratios for the transitions
(\ref{w1}-\ref{w2}) and reduce the errors. Using the present data we find:
\beq{wsum}
BR[D^+ \rightarrow  K^*(892)^+\bar K^o] + 
BR[D^+ \rightarrow  K^*(892)^+\bar K^*(892)^o] = 5.8  \pm 1.9\%
\eeq
This is still large even at two standard deviations. If the large branching
ratios are confirmed with smaller errors, there may be good reason to look for
a new physics explanation.  

\subsubsection{Anomalously high $\eta'$ in charmless strange $B$ decays}.
 
The large experimental branching ratio\cite{PDG} 
$BR(B^+ \rightarrow  K^+ \eta')=6.5\pm 1.7 \times 10^{-5}$ as compared with
$BR(B^+ \rightarrow  K^+ \eta) < 1.4 \times 10^{-5}$ 
and $BR(B^+ \rightarrow  K^o \pi^+)=2.3\pm 1.1 \times 10^{-5}$ 
still has no completely satisfactory explanation and has aroused considerable
controversy\cite{PHAWAII} Also the large inclusive 
$B^+ \rightarrow  K^+ \eta' X$ .branching ratio is equally puzzling.

 A parity selection rule provides a clear experimental method to distinguish
between  two proposed explanations with different flavor topologies . 

1. The OZI-forbidden hairpin diagram\cite{bkpfsi} predicts a  universal
parity-independent enhancement for all  final states arising from the flavor
singlet component of the $\eta'$ \cite{Atwood:1997bn,Halperin:1998ma}..

2.  Parity-dependent interference
between diagrams producing the $\eta'$ via its strange and nonstrange 
components\cite{bkpfsi} predicts a large 
$\eta'$/$\eta$ ratio for even parity final states like $K\eta$ and
$K\eta'$  the reverse for  odd parity states like
$K^*$(892) $\eta$ and $K^*$ $\eta'$\cite{PHAWAII}.
This selection rule agrees with experiment, 
although so far  the $K^*$ $\eta$ has been seen and the $K^*$
$\eta'$ has not.

\subsection{{Predictions from simple easily-tested assumptions}}

\subsubsection{{The Flavor-Topology OZI rule and QCD}}

Two predictions which challenge QCD if they agree with experimenmt.

\beq{nudnik1}
 BR (B^\pm \rightarrow K^\pm \omega) = BR (B^\pm \rightarrow K^\pm \rho^o)
\eeq
Because the $\rho^o$ and $\omega$ mesons both come only from $\overline u u$ 
this prediction requires only exclusion of hairpin diagram topology and
holds even in presence of strong final
state rescattering via all other quark-gluon diagrams.
\beq{nudnik2}
\tilde \Gamma(B^\pm \rightarrow K^\pm \phi)
= \tilde \Gamma(B^\pm \rightarrow K^o \rho^\pm)
\eeq

This prediction also assumes the SU(3) flavor symmetry
relation between strange and nonstrange pair production 

\subsubsection{{The ``inactive spectator" approach}}

Many interesting predictions that can be checked experimentally and challenge
QCD if they work follow from a simple flavor-topology rule\cite{nudat0111}.
The spectator quark line must flow 
continuously from initial to final state, emitting and absorbing gluons
freely, but not undergoing annihilation or pair creation.
One example arises in $B$ decays to
final states containing charmonium. The first prediction 
forbids all  decays without the spectator flavor in the final state :
\beq{nudnika1} 
A[ {B}_d \to J/\psi M(\bar q s)]  =0 =
 A[ {B}_s \to J/\psi M(\bar q d)]  
  \eeq
\beq{nudnika2}
 A( {B_s} \to J/\psi \rho^o) = 
A( {B_s} \to J/\psi \omega) = A( {B}_d \to J/\psi \phi) = 0
\eeq
where $M(\bar q s)$  and $M(\bar q d)$ denote any $\bar q q$ meson with these
constituents.  
If this selection rule holds all other
decays described by two amplitudes  
\beq{nudnika4} 
B(\bar b q) \to J/\psi \bar d q 
\to  J/\psi M(\bar d q) 
\eeq
\beq{nudnika5} 
B(\bar b q) \to J/\psi \bar s q     
     \to
 J/\psi M(\bar s q)  
\eeq

Decay is product of $\bar b \to J/\psi \bar d  ~ \rm{or}  \to J/\psi \bar s $
 decay and hadronization function $h$
\bea
A[ B_d \to J/\psi M^0(\bar s d)]= 
  A( \bar b \to J/\psi \bar s) \cdot  h[\bar s d \to  M^0(\bar s d)] 
 \nonumber \\  
A[ B_s \to J/\psi M^0(\bar d s)] =
  A( \bar b \to J/\psi \bar d) \cdot  h[\bar d s \to  M^0(\bar d s)] 
 \nonumber \\  
  A[ B_d \to J/\psi M^0(\bar d d)] =
 A( \bar b \to J/\psi \bar d) \cdot  h[\bar d d \to  M^0(\bar d d)] 
 \nonumber \\  
 A[ B_s \to J/\psi M^0(\bar s s)] 
  =A( \bar b \to J/\psi \bar s) \cdot  h[\bar s s \to  M^0(\bar s s)]
 \nonumber \\  A[ B^+ \to J/\psi M^+(\bar s u)]
  = A( \bar b \to J/\psi \bar s) \cdot  h[\bar s d \to  M^+(\bar s u)] 
 \nonumber \\  
A[ B^+ \to J/\psi M^+(\bar d u)]
  = A( \bar b \to J/\psi \bar d) \cdot  h[\bar d d \to  M^+(\bar d u)]
\eea

Decays into charge-conjugate strange final states
differ only by weak interaction vertex and kinematic and form  
factor differences induced by   $B_d
- B_s$ mass difference. For any partial wave L in a vector-vector final state,
we can write 

\beq{LCKM1}
 A( {B_s} \to J/\psi  \bar K^{*0})_L = F_{CKM}^{L}\cdot
A( {B_d} \to J/\psi  K^{*0})_L 
\eeq 
If the weak transition is
$
\bar b \to \bar c + W^+ \to \bar c + c + \bar q$
as in the dominant tree diagram.
\beq{LCKM2}
F_{CKM}^{L} =
 {{A_L (\bar b \to J/\psi  \bar d)}\over
{A_L( \bar b \to J/\psi \bar s)}} \approx V_{cd}/V_{cs}
\eeq
 
If $ F_{CKM}^L \not=  V_{cd}/V_{cs}$ other contributions are indicated.
 
Additional SU(3) assumption gives 
\beq{LCKM5}
A_L( {B_d} \to J/\psi \rho^o) 
  = A_L( {B_d} \to J/\psi \omega) = 
 {{A_L( {B_s} \to J/\psi  \bar K^{*0})}\over {\sqrt 2}}  
\eeq 
\beq{LCKM6}
A_L( {B}_s \to J/\psi \phi) 
  = A_L( {B}_d \to J/\psi  K^{*0})
\eeq 
\beq{LCKM7}
A_L( {B_s} \to J/\psi \rho^o) 
  = A_L( {B_s} \to J/\psi \omega) = A_L( {B}_d \to J/\psi \phi) = 0
\eeq

\subsubsection{How to test $\eta-\eta'$  mixing }

 If the $\eta-\eta'$  system satisfies the standard mixing,

\beq{mixing} \ket {\eta}  =  \ket {\eta_n} \cos \phi -  \ket {\eta_s} \sin \phi;  $$
~  ~ ~  \ket {\eta'} =  \ket {\eta_n} \sin \phi +  \ket {\eta_s} \cos \phi 
\eeq

Then
\beq{mixtexst1}
  r_d = {{
{p_{\eta^{\prime}}^3\Gamma( {\bar B}^0 \rightarrow J/\psi \eta) }}\over{
{p_{\eta}^3\Gamma( {\bar B}^0 \rightarrow J/\psi \eta^{\prime})} }}= 
\cot^2\phi; ~ ~ ~  
r_s= {{{p_{\eta^{\prime}}^3\Gamma( {\bar B_s} \rightarrow J/\psi \eta) }}\over{
{p_{\eta}^3\Gamma( {\bar B_s} \rightarrow  J/\psi \eta^{\prime})} }}
= \tan^2\phi
\eeq
\beq{mixtexst2}
R\eta = {{ 
{p_{B\eta}^3\Gamma( {\bar B}^0 \rightarrow  J/\psi \eta) }}\over{
{p_{Bs\eta}^3\Gamma( {\bar B_s} \rightarrow  J/\psi \eta)}}} \cdot
{{
{p_{BsK}^3\Gamma( {\bar B_s} \rightarrow  J/\psi  K^0)}}\over{
{p_{BK}^3\Gamma( {\bar B}^0 \rightarrow  J/\psi \bar K^0) }
}}= \cot^2\phi
\eeq
\beq{mixtexst3}
R_\eta^{\prime} = {{
{p_{B\eta^{\prime}}^3\Gamma( {\bar B}^0 \rightarrow  J/\psi \eta^{\prime}) }
 }\over{
{p_{Bs\eta^{\prime}}^3\Gamma( {\bar B_s} \rightarrow  J/\psi \eta^{\prime})} 
}}\cdot
{{
{p_{BsK}^3\Gamma( {\bar B_s} \rightarrow  J/\psi  K^0)}}\over{
{p_{BK}^3\Gamma( {\bar B}^0 \rightarrow  J/\psi \bar K^0) }
}}= 
\tan^2\phi
\eeq
\beq{mixtexst4}
r  = \sqrt{r_dr_s}=1; ~ ~ ~ R_B  =  \sqrt{R_\eta R_\eta^{\prime}}=1
\eeq  
Any large deviation of $r$ or $ R_B$  from 1 would indicate evidence of non
standard $\eta-\eta^{\prime}$ mixing\cite{nudat0111}.

\subsubsection {SU(3) Relations between Cabibbo-Favored and Doubly-Cabibbo Suppressed
$D^o$ decays}.

The SU(3) transformation $d \leftrightarrow s $, also called a  Weyl reflection
or a   U-spin reflection relates Cabibbo-favored $\leftrightarrow $
doubly-cabibbo suppressed charm decays 
\bea     
d \leftrightarrow s  ; ~ ~ ~    K^+ \leftrightarrow \pi^+  ; ~ ~ ~
  K^- \leftrightarrow \pi^-  ;  ~ ~ ~  
    D^+ \leftrightarrow D_s; \nonumber \\  
 D^o \leftrightarrow D^o;  ~ ~ ~ 
   K^+ \pi^- \leftrightarrow  K^- \pi^+         
\eea

If strong interaction final state interactions
conserve SU(3) the only SU(3) breaking occurs in the CKM matrix elements.

A simple  test  of this SU(3) symmetry is
\beq{nudnika} 
 tan^4 \theta_c = {{BR( D^o \rightarrow  K^+ \pi^-)}\over{BR(D^o \rightarrow  
K^-\pi^+) }} =
{{BR[ D^o \rightarrow  K^*(892)^+\rho^-]}\over{BR[D^o \rightarrow  
K^*(892)^-\rho^+]}} 
\eeq 
These relations involve only branching ratios and are easily tested . 
 They involve  no phases and only
branching ratios of decay modes all expected to be comparable to the observed
DCSD $D^o \rightarrow K^+ \pi^-$. 
A similar relation . 
\beq{nudnikb} 
tan^4 \theta_c = 
 {{BR[ D^o \rightarrow  K^+  a_1(1260)^-]}\over{BR[D^o \rightarrow  
K^- a_1(1260)^+] }} 
\eeq
may be subject to a different type of SU(3) breaking.
. A weak vector dominance form factor can enhance 
\beq{nudnikc}
D^o(c\bar u)
\rightarrow  (s \bar u\rightarrow K^{-} )_S \cdot 
(u\bar d\rightarrow  a_1^+ )_W
\rightarrow  K^{-} a_1^+ 
\eeq
where the subscripts S and W  denote strong and weak form factors. 

No such enhancement should occur in 
\beq{nudnikd}
D^o(c\bar u) 
\rightarrow  (d \bar u \rightarrow a_1^-)_S \cdot 
(u\bar s \rightarrow K^{+} )_W
\rightarrow  a_1^-  K^{+ } 
\eeq

If the SU(3) breaking is really due to the difference between products
of weak axial and strong kaon form factors and vice versa, the SU(3) relation 
invlving the $a_1$ can be expected to be strongly broken and replaced by the 
inequality
\beq{nudnike}
{{BR[D^o \rightarrow  
K^- a_1(1260)^+]}\over{BR(D^o \rightarrow  
K^-\pi^+) }}
 \gg {{BR[ D^o \rightarrow  K^+  a_1(1260)^-]}\over{
BR( D^o \rightarrow  K^+ \pi^-) }} 
\eeq 

\subsubsection {A problem with strong phases}.
 
The $d \leftrightarrow s$ interchange SU(3) 
transformation also predicts\cite{lincoln}
  $ D^o \rightarrow K^+ \pi^-$ and $ D^o \rightarrow K^- \pi^+ $  
have the same strong phases. 
This has been shown 
to be in disagreement with experiment\cite{sven} showing SU(3) violation.  
   
But the $ K^+
\pi^- $ and $ K^- \pi^+ $ final states are charge conjugates of one another
and strong interactions conserve charge conjugation.
SU(3) can be
broken in strong interactions without breaking charge conjugation only in the
quark - hadron form factors arising in hadronization transitions like  
 \bea 
D^o(c\bar u) &
\rightarrow  (s \bar u\rightarrow K^{-} )_S
\cdot  (u\bar d\rightarrow  a_1^+ )_W  \nonumber \\ 
& \rightarrow  K^{-} a_1^+  \rightarrow
K^-\pi^+ 
\eea 
 \bea D^o(c\bar u) & 
\rightarrow  (d \bar u \rightarrow a_1^-)_S \cdot (u\bar s \rightarrow K^{*+} )_W
 \nonumber \\ 
&\rightarrow  a_1^-  K^{+}  \rightarrow \pi^- K^+ 
\eea  
with the SU(3)  breaking given by the inequality (\ref{nudnike}).

The $a_1$ and $\pi$  wave functions are very different and  not related by
SU(3). 
The $ K^{\mp} a_1^\pm \rightarrow K^\mp\pi^\pm$ transition
can proceed via $\rho$ exchange 

\subsubsection {SU(3) relations between $D^+$ and $D_s$ decays}

Both of the following ratios of branching ratios 
 \beq{QQ1a}     
{{BR( D_s \rightarrow  K^+K^+\pi^-)}\over{BR(D_s \rightarrow  
K^+K^-\pi^+)}} \approx {{BR( D^+ \rightarrow 
 K^+\pi^+\pi^-)}\over{BR(D^+\rightarrow  
K^-\pi^+\pi^+) }}  
\approx O(tan^4 \theta_c)   
\eeq 
are ratios of a doubly Cabibbo forbidden decay to an allowed
decay and should be of order $tan^4 \theta_c$. The SU(3) transformation $d
\leftrightarrow s$ takes the two ratios (\ref{QQ1a}) into the
reciprocals of one another. SU(3) requires
the product of these two ratios to be EXACTLY $\tan^8 \theta_c$\cite{plipzh}.
 \beq{tan}
{{BR( D_s \rightarrow  K^+K^+\pi^-)}\over{BR(D_s \rightarrow K^+K^-\pi^+)}}
\cdot {{BR( D^+ \rightarrow  K^+\pi^+\pi^-)}\over{BR(D^+\rightarrow
K^-\pi^+\pi^+) }} = tan^8 \theta_c
\eeq  

Most  obvious
SU(3)-symmetry-breaking factors cancel out in this product; e.g. phase space.
Present data\cite{PDG} show
\beq{dub1}
 {{BR(D^+\rightarrow K^+\pi^-\pi^+)}\over{BR(D^+ \rightarrow K^-\pi^+\pi^+)}}
  \approx
  0.65\% \approx 3 \times \ tan^4 \theta_c
\eeq
Then SU(3) predicts
\beq{dub2}
 {{BR(D_s\rightarrow K^+K^+\pi^-)}\over{BR(D_s\rightarrow K^+K^-\pi^+)}}
\approx
{{tan^4 \theta_c}\over{3}} \approx  0.07\%.
\eeq  

If this SU(3) prediction is confirmed
experimentally some new dynamical explanation will be needed for the 
order of magnitude difference between effects of the final-state
interactions in $D^+$ and $D_s$ decays. 

If the final state interactions  behave similarly 
in $D_s$ and $D^+$ decays,
the large violation of SU(3) will need some explanation.

New physics enhancing the doubly suppressed decays
might produce a CP violation  observable as a charge asymmetry in the products
of above the two ratios; i.e between the values for $D^+$ and $D_s$ decays and
for $D^-$ and $\bar D_s$ decays.

An obvious caveat is the almost trivial SU(3) breaking arising
from resonances in the final states. But  sufficient data and Dalitz plots
should enable including these effects.
In any case the SU(3) relation and its possible violations raise interesting
questions which deserve further theoretical and experimental investigation.
Any really large SU(3)-breaking final state interactions
that we don't understand must cast serious doubts on many SU(3)
predictions.

\section*{Acknowledgments}

 This work was supported
in part by the U.S. Department
of Energy, Division of High Energy Physics,  
Contract W-31-109-ENG-38.and
in part by a grant from the United States-Israel
Binational Science Foundation (BSF), Jerusalem, Israel and
by the Basic Research Foundation administered by the Israel Academy of 
Sciences and Humanities

\end{document}